\documentclass{ws-ijmpcs}

\begin{document}

\markboth{Naso L., Miller J. and Klu{\'z}niak W.}
{Magnetic fields in accretion disks around neutron stars}

\catchline{}{}{}{}{}
%
%%%%%%%%%%%%%%%%%%%%%%%%%%%%%%%%%%%%%%%%%%%%%%%%%%%%

\title{MAGNETIC FIELDS IN ACCRETION DISCS AROUND NEUTRON STARS --
CONSEQUENCES FOR THE CHANGE OF SPIN}

\author{LUCA NASO}

\address{Key Laboratory of Solar Activity, National Astronomical Observatories, 
Chinese Academy of Sciences,
20A Datun Road, Chaoyang District, Beijing 100012, China -
luca.naso@gmail.com}

\author{JOHN MILLER  and  WLODEK KLU{\'Z}NIAK}

\address{Department of Physics (Astrophysics), University of Oxford, Keble
Road, Oxford OX1 3RH, UK\\
Nicolaus Copernicus Astronomical Center, Polish Academy of Sciences, 
ul. Bartycka, 18, PL-00-716 Warszawa, Poland}

\maketitle

\begin{history}
\received{28 December 2012}
%\revised{Day Month Year}
%published <= 23 November 2013
\end{history}

\begin{abstract}
Accretion disks are ubiquitous in the universe and it is generally accepted that
magnetic fields play a pivotal role in accretion-disk physics. The spin history
of millisecond pulsars, which are usually classified as magnetized neutron stars
spun up by an accretion disk, depends sensitively on the magnetic field
structure, and yet highly idealized models from the 80s are still being used for
calculating the magnetic field components.
We present a possible way of improving the currently used models with a
semi-analytic approach. The resulting magnetic field profile of both the
poloidal and the toroidal component can be very different from the one suggested
previously. This might dramatically change our picture of which parts of the
disk tend to spin the star up or down.

\end{abstract}

\ccode{PACS numbers: 97.10.Gz, 97.60.Jd, 97.60.Gb}

\section{Introduction}
Neutron stars (NSs) in binary systems can constantly attract matter from
their companions. Depending on its angular momentum, matter can reach the NS
through a spherically symmetric wind, or can instead create a disk-like
structure around the star. If there were no viscosity in the disk, it 
could be in perfect Keplerian rotation and no matter would accrete onto the NS.
However what we see from observations is that accretion does occur, therefore
there must be some kind of viscosity in the disk which is responsible for
transporting angular momentum outward and matter inward. Molecular viscosity
proved to be too small to justify the required accretion rates. The best
possible source for this viscosity is turbulence, which could be generated by
the magnetic field through the magneto-rotational instability.

In addition to generating the turbulence, magnetic fields can have many other
effects on the whole system, such as modifying the spectral properties and
launching jets or winds. We are here interested in the spin history of the
central object and we consider the recycled pulsar scenario, where an old NS is
spun up by an accretion disk to become a millisecond pulsar. Within this
scenario the magnetic field plays a very important role because it creates a
strong link between part of the disk and the star. Therefore the whole
disk contributes to angular momentum exchange with the NS and not only the
accreted matter in the inner disk region. Moreover the magnetic torque can have
both signs, i.e. it can both spin the star up or down.

Currently, simplified models from the 80s are used to calculate both the
magnetic field structure and the related magnetic torque. Our goal is to
improve on those models by following a step-by-step procedure, which can also
help us in deconstructing what we see from large numerical simulations.

\section{Current Models}
Although large numerical calculations are presently available, simplified
analytic models developed in the 80s are still being used. These take the
poloidal component of the magnetic field $B_{\rm pol}$ to be a perfect dipole,
the disk to be thin and neglect all radial derivatives with respect to vertical
ones. Moreover disk matter is assumed to have only toroidal motion with
Keplerian angular velocity. By further making the kinematic approximation, i.e.
no feedback from the magnetic field on the velocity field, the following
analytic solution is found (Wang\cite{W87} and Campbell\cite{C87}):
\begin{equation}
\label{eq:bp_an}
B_\phi = \gamma_{\rm a} \, (\Omega_{\rm disk} - \Omega_{\rm s}) \, B_z
\, \tau_{\rm d} \, \propto \, \Delta \Omega / \varpi^3 \, {\rm ,}
\end{equation}
where $\varpi$ is the cylindrical radius; $\gamma_{\rm a}$ is the amplification
factor (which is is of the order of unity and depends on the steepness of the
vertical transition from Keplerian rotation inside the disk to corotation with
the star at the top of the disk); $\Omega_{\rm disk}$ and $\Omega_{\rm s}$ are
the disk and stellar angular velocities, respectively; and $\tau_{\rm d}$ is
the dissipation time scale. 
Equation~(\ref{eq:bp_an}) states that the disk rotation distorts the original
magnetic field and creates a toroidal component, $B_\phi$, which is proportional
both to $B_z$ and to the relative angular velocity $\Delta\Omega$ between the
disk and the central star.

Once the magnetic field structure is known, one can calculate the magnetic
torque exerted by the disk on the star. This is usually done following Ghosh
and Lamb\cite{GL79a}$^{,}$\cite{GL79b} who, by making the same assumptions as
for Eq.~(\ref{eq:bp_an}), obtained:
\begin{equation}
 %\label{eq:tor_an}
 T_{B} \propto B_\phi \, B_z \propto \Delta \Omega \, B_z^2
\end{equation}
 This equation states that (i) the inner part of the disk (i.e.
$r<r_{\rm cor}$\footnote{The corotation radius $r_{\rm cor}$ is defined as the
radius where a Keplerian disk rotates with the same angular velocity as the
central star.}) provides a spin-up torque  and that (ii) the outer part of
the disk instead gives spin-down.

Many investigators have simulated the magnetosphere--disk interaction, solving
the full set of the MHD equations (see e.g. Romanova et al.\cite{Ral11}). The
work that we are  presenting here should not be seen as being in competition
with these  analyses, but rather as being complementary to them, by providing
both a useful test case and a means for getting a better understanding of the
underlying processes.

\section{Our strategy}
The conceptual papers from the 80s mentioned above, continue to be widely quoted
and used as the basis for new research (see e.g. Kluzniak and
Rappaport\cite{KR07}) and our work here stands in the tradition of refining
these approaches. Our strategy can be summarized with the following four steps:
(i) begin by defining an appropriate simple model; (ii) obtain a reduced set of
equations for the model; (iii) solve for the magnetic field structure; (iv)
solve for the torque exerted on the neutron star.

\subsection{Initial model}
Our initial model represents a rotating NS, with a dipole magnetic field aligned
with the rotation axis. The disk is truncated at the Alfven radius and has a
corona above and below it, treated as a boundary layer between the disk and
surrounding vacuum. We impose dipole boundary conditions (BCs) on all
boundaries, assume axisymmetry ($\partial_\phi[\dots]=0$), stationarity
($\partial_t[\dots]=0$) and neglect the magnetic feedback.

The model is fully 2D (no vertical integration or Taylor expansion) and includes
all components of $\vec{v}$. We make no assumption about the magnetic field
components and solve self-consistently for both the poloidal and toroidal
components.

For the velocity field in our first model\cite{papI}$^{,}$\cite{papII}
we used the Shakura and Sunyaev profile\cite{SS73} (SS), suitably modified to
make it consistent with dipolar BCs and fully 2D. In our second
model\cite{papIII} we used a more general 2D profile, with a backflow in the
equatorial region (Kluzniak and Kita\cite{KK00}, KK).

\subsection{Equations}
We consider the induction equation for mean fields: 
$
\partial_t \vec{B} = \nabla\times\left(  \vec{v} \times \vec{B}  - \eta_{\rm T}
\nabla \times \vec{B} \right)
$, where $\eta_{\rm T}$ is the turbulent magnetic diffusivity and dynamo effects
are neglected. We then write this equation in spherical coordinates and apply
all of the hypotheses mentioned earlier. This gives the following set of
equations:
\begin{align}
\label{eq:ind_r}
0 =& \, \partial_\theta \left\{ \sin\theta \left[ v_r B_\theta - v_\theta 
B_r - \frac{\eta_{\rm T}}{r}[\partial_r(rB_\theta) - \partial_\theta
B_r]\right] 
\right\} \\
\label{eq:ind_t}
0 =& \, \partial_r \left\{ r \hspace{0.6cm}\left[ v_r B_\theta - v_\theta 
B_r - \frac{\eta_{\rm T}}{r}[\partial_r(rB_\theta) - \partial_\theta B_r]
\right] 
\right\} \\
\label{eq:ind_p}
\nonumber
0 =& \, \partial_r \left\{ r \left[ v_\phi B_r - v_rB_\phi + \frac{\eta_{\rm
T}}{r} \partial_r(rB_\phi) \right] \right\} - \\
&\,\partial_\theta \left\{ v_\theta B_\phi - v_\phi B_\theta  -  
\frac{\eta_{\rm T}}{r\sin\theta}\partial_\theta (B_\phi \sin\theta) \right\} 
\,\, \mbox{.}
\end{align}
There is the partial decoupling in this set of equations, i.e. the equations
describing the poloidal structure do not contain any azimuthal quantity. We can
therefore solve Eqs.~(\ref{eq:ind_r}) and (\ref{eq:ind_t}) first and then use
the results to solve Eq.~(\ref{eq:ind_p}).

We solve this set of partial differential equations numerically, by using the
Gauss-Seidel relaxation method on a grid whose resolution is $\Delta
r\sim0.7\,r_g$ and $\Delta\theta\sim0.1^{\circ}$, where $r_{\rm g}\sim4$ km is
the Schwarzschild radius for a $1.4\,M_\odot$ NS.

\subsection{Magnetic field structure}
\begin{figure}[ht]
$\begin{array}{lll}
\psfig{file=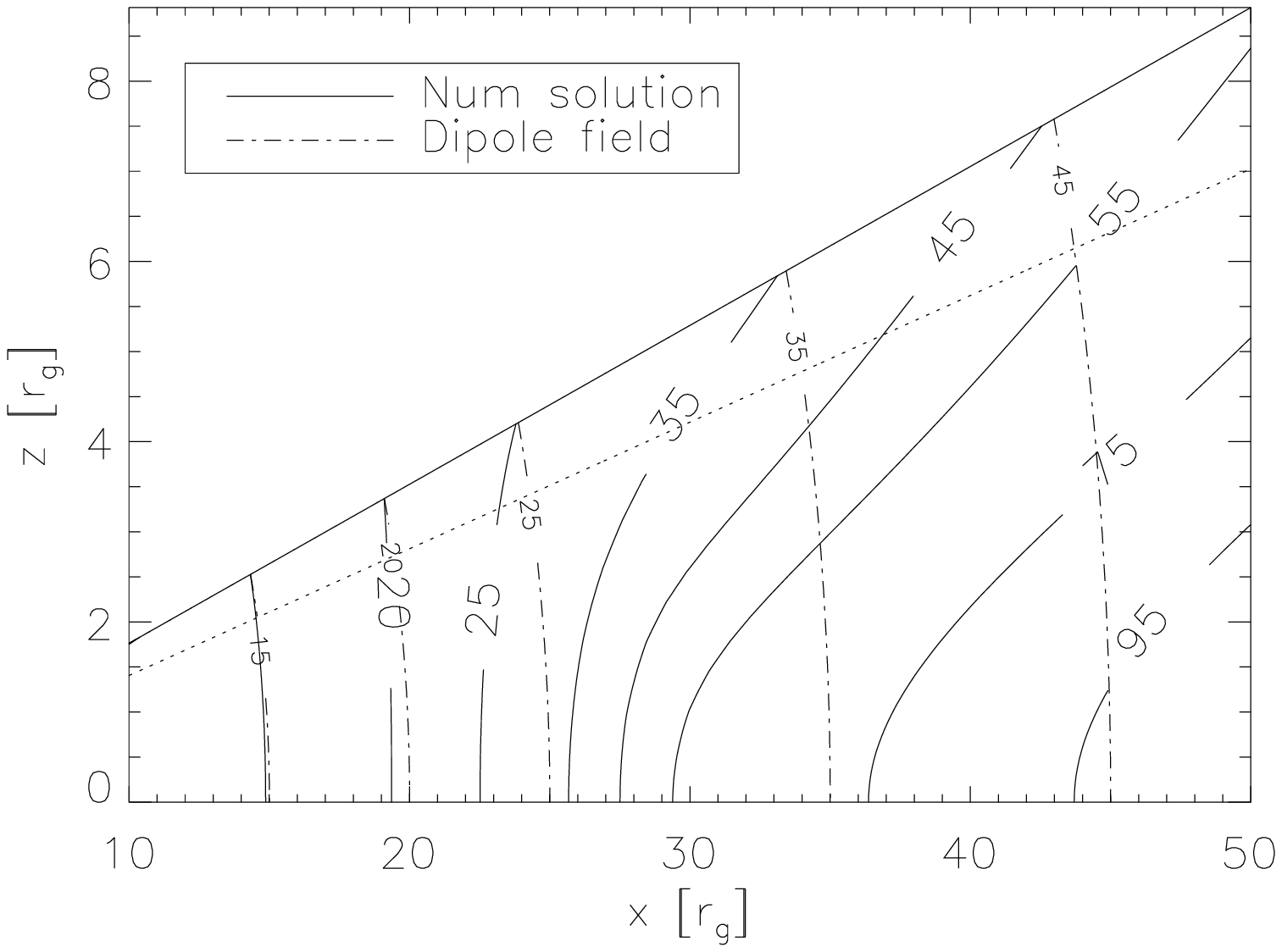,width=0.45\textwidth} & \hspace{0.8cm} &
\psfig{file=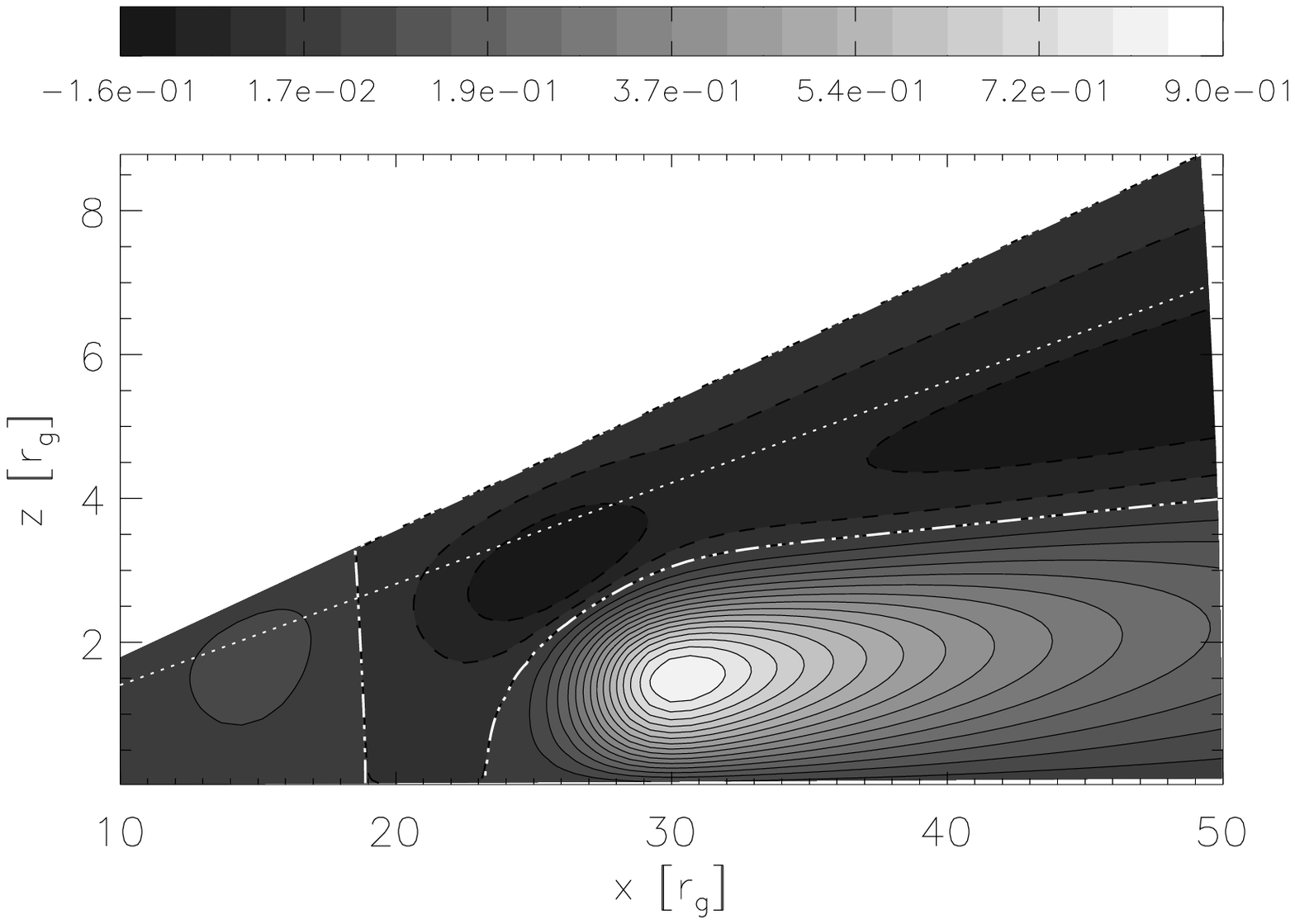,width=0.45\textwidth}
\end{array}$
\vspace*{4pt}
\caption{Magnetic field structure for our first model\protect\cite{papI}$^{,}$\protect\cite{papII} with SS velocity field. The dotted line marks the boundary between the disk and the corona. Left panel: field lines in a vertical plane. Right panel: contour plot of $B_\phi$. Solid contours indicate positive values, while dashed lines are used for negative ones. Triple-dotted-dashed lines are used to mark zero level.\label{fig:mf}}
\end{figure}

\begin{figure}[ht]
$\begin{array}{lll}
\psfig{file=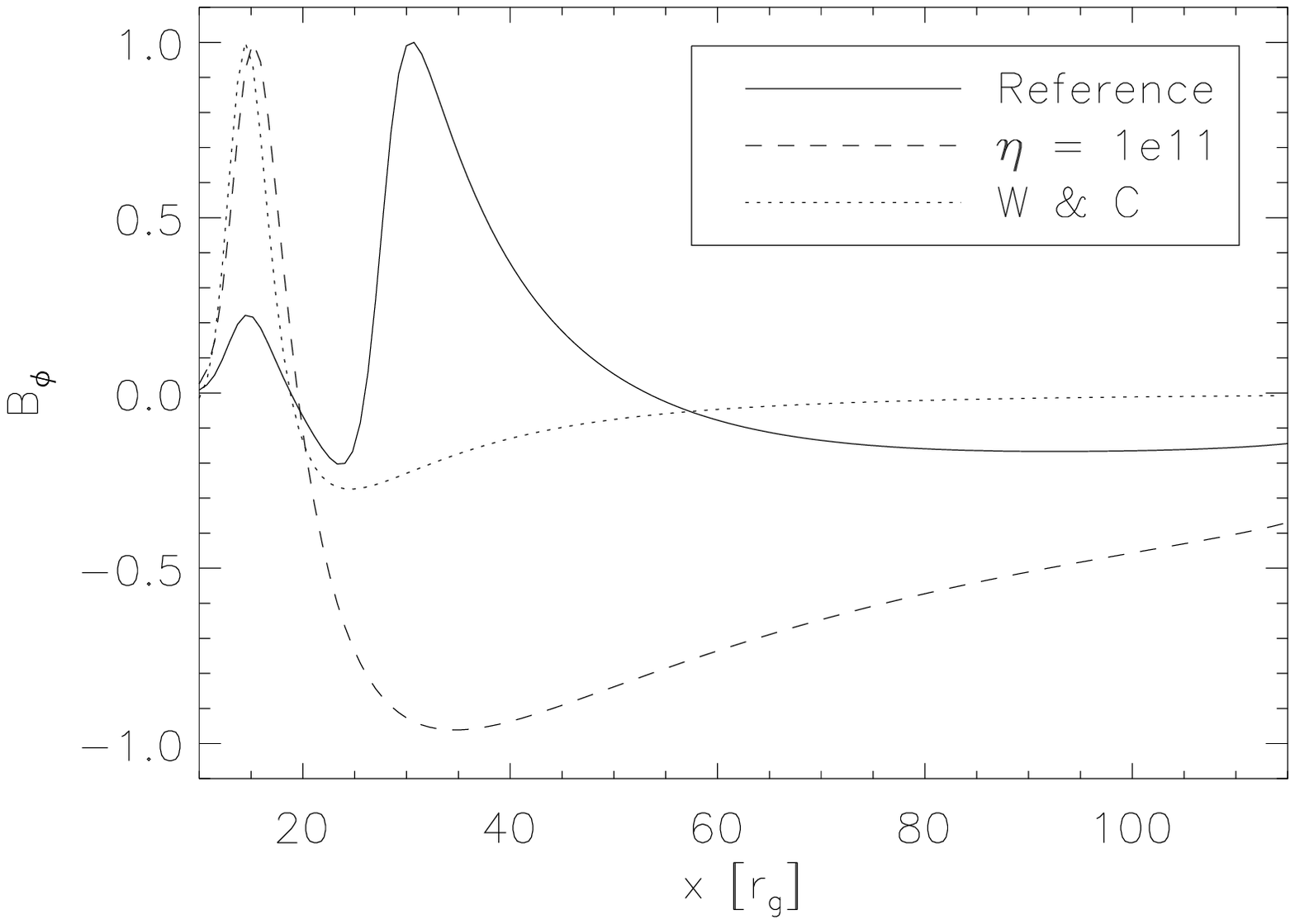,width=0.45\textwidth} & \hspace{0.8cm} &
\psfig{file=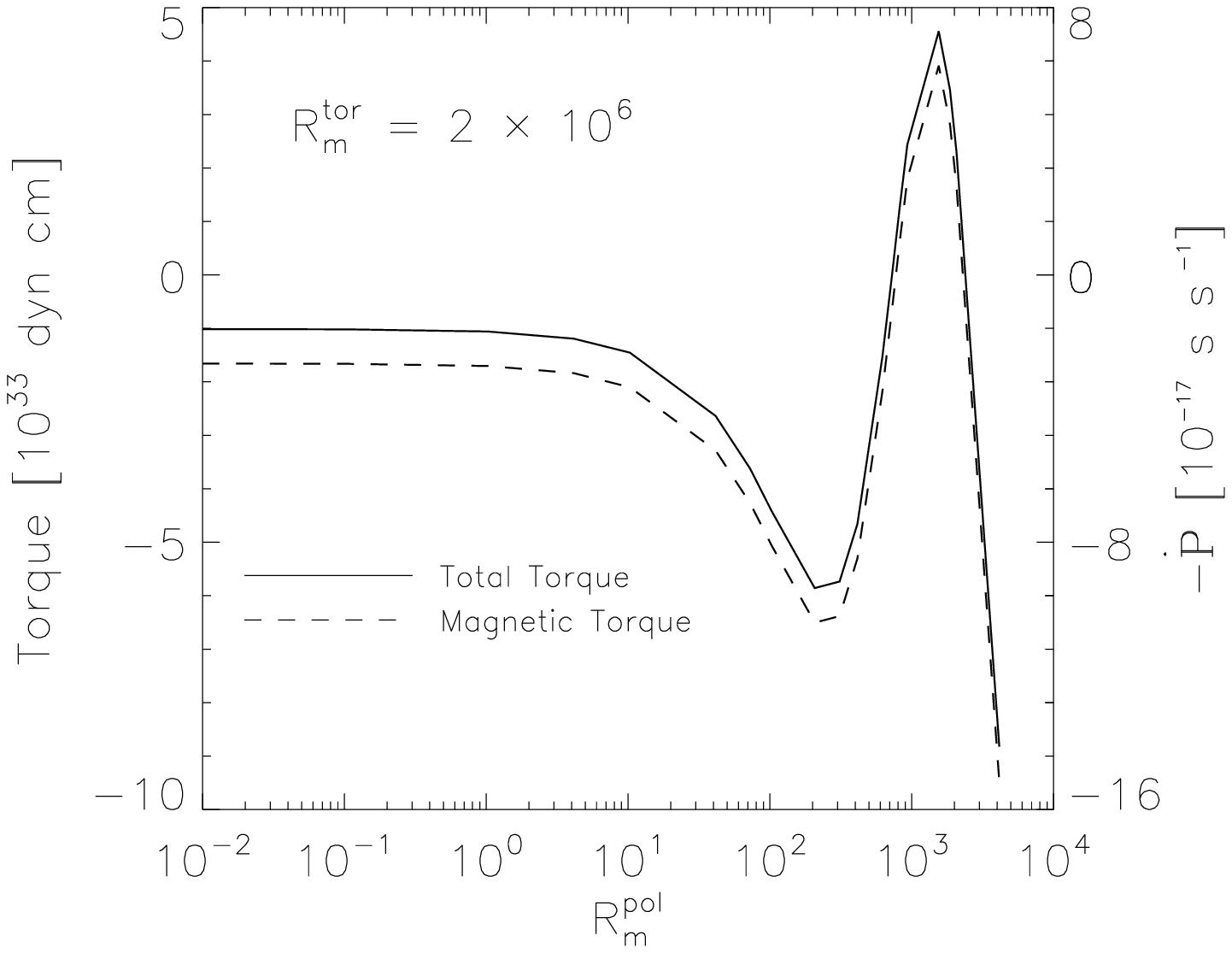,width=0.45\textwidth}
\end{array}$
\vspace*{4pt}
\caption{Left panel: Shell averages of $B_\phi$. The solid and dashed lines are two configurations from our numerical analysis for our first model\protect\cite{papI}$^{,}$\protect\cite{papII}, the dotted line is the result of Wang\protect\cite{W87} and Campbell\protect\cite{C87}. All of the curves are normalized to 1. Right panel: magnetic torque for our second model\protect\cite{papIII} with KK velocity field, plotted against $R_m^{\rm pol}$, for constant $R_m^{\rm tor}$.\label{fig:tor}}
\end{figure}

In the previous simplified models the poloidal component of the magnetic field
was taken to be a pure dipole, however, we find that deformations of $B_{\rm
pol}$ can be quite substantial as a result of the poloidal motion of the matter.
The left panel of Fig.~\ref{fig:mf} shows a plot of the magnetic field lines in
a vertical section through the disk. The toroidal field can also be very
different from the predictions of Eq.~(\ref{eq:bp_an}). As shown in the right
panel of Fig.~\ref{fig:mf}, it has a fully 2D structure with multiple zeros
which are not only at the corotation radius. To facilitate comparison, in the
left panel of Fig.~\ref{fig:tor} we plot the shell average of $B_\phi$ for two
of our models, and as given by Eq.~(\ref{eq:bp_an}).

%
%\begin{figure}
%$\begin{array}{lll}
% \includegraphics[width=0.45\textwidth]{nasoluca_fg1a.eps} & \hspace{0.8cm} & \includegraphics[width=0.45\textwidth]{nasoluca_fg1b.eps}
% \end{array}$
%  \caption{Magnetic field structure for our first model\cite{papI}$^{,}$\cite{papII} with SS velocity field. The dotted line marks the boundary between the disk and the corona. Left panel: field lines in a vertical plane. Right panel: contour plot of $B_\phi$. Solid contours indicate positive values, while dashed lines are used for negative ones. Triple-dotted-dashed lines are used to mark zero level.}
% \label{fig:mf}
%\end{figure}
%

%

\subsection{Torque}
The torque exerted on the NS by the disk is the moment of the Lorentz force:
\begin{equation}
\label{eq:tor}
 \vec{T}_{B-{\rm disk}} = - \int\limits_{\rm disk} \vec{r} \times
\vec{F}_{\rm L} \, dV \,\, \mbox{.}
\end{equation}
We calculate $\vec{T}_{B-{\rm disk}}$ without neglecting radial derivatives and
using our previous results for the magnetic field. We studied $\sim30$
configurations and found that, even with the same values of the NS period and
magnetic field, the rate of period change $\dot{P}$ can be positive, negative or
zero depending on value of the poloidal magnetic Reynolds number. We find $\dot
P$ to be in the range:
\begin{align}
- \dot{P} =  [-5.3,\, +4.5] \times 10^{-16}\,\, [{\rm s/s}] \,\, \mbox{,}
\end{align}
while in the Wang model\cite{W87} $- \dot{P}_{\rm W87} = 3 \times
10^{-17}\, [{\rm s/s}]$. For comparison, the infalling accreting matter alone
would give $- \dot{P}_{\rm matter} = 10^{-17} \, [{\rm s/s}]$.

\section{Conclusions}
Simplified models from the 80s are still being used for describing magnetic
fields in accretion disks. We have studied more realistic models and found very
different results. $B_{\rm pol}$ shows large deviations away from the dipole
that is assumed in the previous models: the field lines are pushed inward and
accumulate in the inner part. Also $B_\phi$ is different: it is not proportional
to $\Delta\Omega$ and shows a fully 2D structure. As regards the magnetic
torque: this can dominate over the matter torque, as in the simplified models.
However, it can vary substantially depending on the velocity field and turbulent
diffusivity in the disk, even if the magnetic field and spin period of the
central neutron star are kept fixed.

\end{document}